\documentclass[12pt]{article} 

\usepackage[pdftex]{graphicx}
\usepackage{url}

\begin{document}

\title{Non-Blocking Signature of very large SOAP Messages}
\author{G.A. Kohring and L. Lo Iacono\\
IT Research Division, NEC Laboratories Europe\\
Rathausallee 10, D-53757 Sankt Augustin, Germany\\
kohring@it.neclab.eu}
\date{September 17, 2007}

\maketitle

%%%%%%%%%%%%%%%%%%%%%%%%%%%%%%%%%%%%%%%%%%%%%%%%%%%%%%%%%%%%%%%%%%%%%%%%%%%%%%
\begin{abstract}
Data transfer and staging services are common components in Grid-based, or
more generally, in service-oriented applications. Security mechanisms
play a central role in such services, especially when they are deployed in 
sensitive application fields like e-health. The adoption of WS-Security and 
related standards to SOAP-based transfer services is, however, problematic as
a straightforward adoption of SOAP with MTOM introduces considerable 
inefficiencies in the signature generation process when large data sets are 
involved. This paper proposes a
non-blocking, signature generation approach enabling a stream-like processing
with considerable performance enhancements.

\end{abstract}

%%%%%%%%%%%%%%%%%%%%%%%%%%%%%%%%%%%%%%%%%%%%%%%%%%%%%%%%%%%%%%%%%%%%%%%%%%%%%%
\section{Introduction}
Many Grid-based applications require data transfer and staging services in 
order to deliver input data to and output data from compute services. Depending
on the concrete application field, security policies play a vital role in 
such services and are often a critical distinguishing factor.

In medical treatment or research scenarios, in which medical images are 
transferred to simulation services, the confidentiality, integrity and 
authenticity of the image data as well as the returned simulation results 
have to be ensured \cite{ehealth}. An according end-to-end communication 
security component 
is a necessary building block for a secure transfer service for such 
environments.

\begin{figure}[htp]
 \includegraphics[width=4.6in]{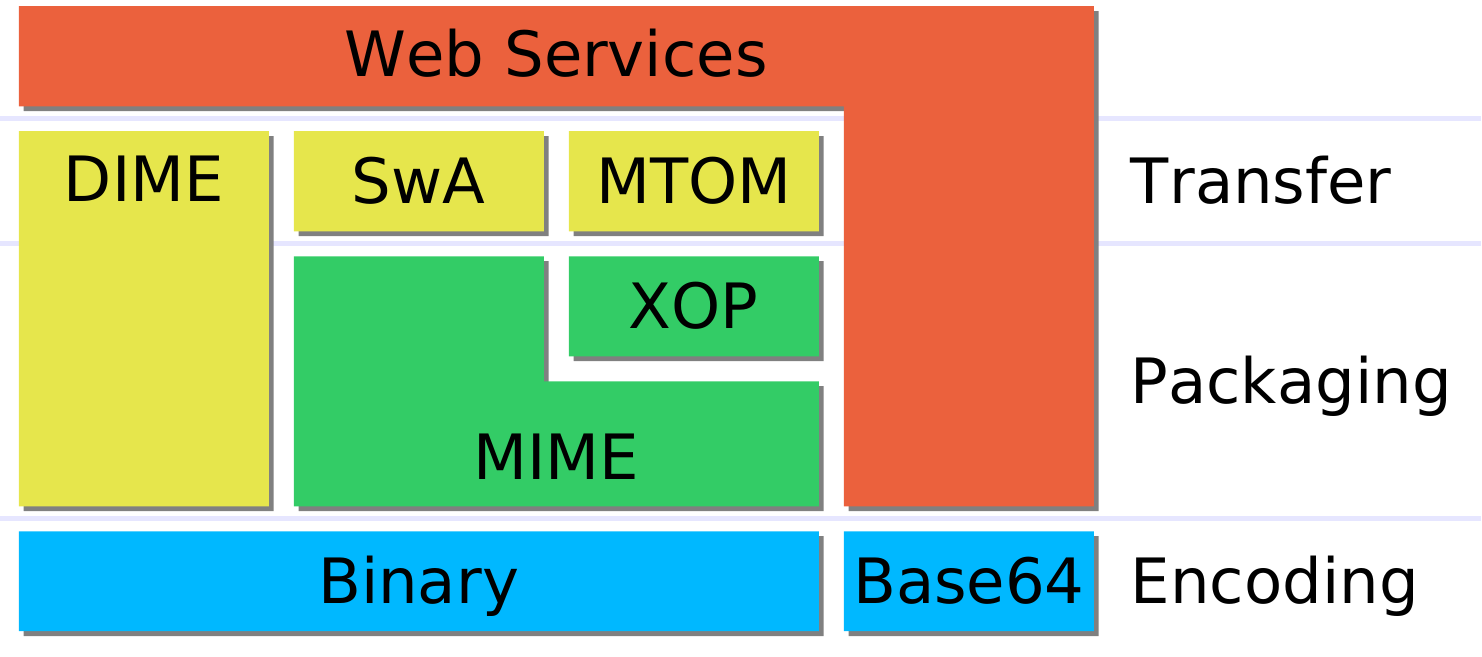}
 \caption{SOAP Data Transfer Protocol Stack}
 \label{fig:datatransfsoap}
\end{figure}

Due to the fact that Grids are more and more converging to Web service 
technologies and accompanying standards, the application of WS-Security and related specifications seems to be an obvious solution to provide such security mechanisms for data transfer services. A closer look at the available technologies for data transfer using SOAP reveals, however, that it is not as straightforward as expected. Fig.~\ref{fig:datatransfsoap} provides an overview of the 
available technologies and their relationship for transferring (binary) data 
with SOAP.

Since the SOAP protocol elements are XML-encoded, data transfer using SOAP falls back to embedding the data into XML documents. XML is usually presented as a way of describing text data within the context of a well-formed document containing meta-information (which is also text based) meant to bring some structure and form to the text data. There are, however, various domains that do not lend themselves nicely to being represented with textual data only. Thus, technologies for the inclusion of binary data into XML documents are needed and are obviously playing an important role in data transfer.

There are several approaches to circumvent the binary inclusion problem. A
common approach is to encode the binary data into some string representation.
In fact, the XML Schema \cite{XMLSchemaPart2} defines the base64\-Binary 
type that can be used for this purpose. 3 octets of binary data are mapped to 4 octets of base64-encoded data introducing a data expansion of 33\% for UTF-8 text encoding (for UTF-16 text encoding the data expansion will double) as well as additional processing costs for coding and decoding.

SOAP with Attachments (SwA) \cite{SwA} is a W3C recommendation defining a way for binding attachments to a SOAP envelope using the multipart/related MIME type. The binary data is in an MIME attachment. It is referred to from the SOAP message with a \texttt{cid:}-URI, which uses the value of the Content-ID MIME header to find the corresponding attachment. This combination of URI reference and raw data inclusion avoids the overhead and bloat of encoding, but introduces other limitations. MIME uses text strings to delineate boundaries between attachment parts. Consumers must scan the entire message to find the string value used to delineate a boundary. Due to the avoidance of an explicit length field, however, the MIME specification places no actual limit on the size of attachments. MIME cannot be represented as an XML Infoset which effectively breaks the Web services model causing e.g that attachments cannot be secured using WS-Security \cite{WSSec} in a straightforward way.

A seperate profile has been published, which explicitly governs the usage of
WS-Security with SwA \cite{WSSec-SwA}. More specifically, it describes 
securing SOAP attachments using SOAP message security for attachment integrity,
confidentiality and origin authentication, and receiving process of such a message. Furthermore, the standard allows the choice of securing MIME header information exposed to the SOAP layer, and also allows MIME transfer encodings to be changed to support MIME transfer, despite support for integrity protection and SwA messages to transit SOAP intermediaries. However, a choice of transport layer security (e.g. SSL/TLS), S/MIME, application using XML Signature and XML Encryption, and other SOAP attachment mechanisms (e.g. MTOM) is explicitly out of scope of this standard, and persisting signatures and signing portions of attachments are not considered neither. It furthermore needs to be taken into account, that before the security transformation can be performed, the attachment needs to be canonicalized according to its MIME type. Thus, various MIME type specific canonicalizations need to be supported, when applying this approach to secure SOAP attachments.

The Direct Internet Message Encapsulation (DIME) \cite{DIME} is a Microsoft
supported Internet standard for the transfer of binary and other encapsulated 
data over SOAP. The standard can be seen as an alternative to SwA and was supposed to be a simplified and more efficient (in terms of decoding time) version of MIME. The initial draft was submitted to the IETF in November 2001. The last update was submitted in June 2002. By December 2003, DIME had lost out, in competition with Message Transmission Optimization Mechanism (MTOM, see below) \cite{MTOM} and SwA, and Microsoft now describes it as ``\emph{superseded by the SOAP Message Transmission Optimization Mechanism (MTOM) specification}''. The DIME specification was created to address performance issues when processing MIME attachments. DIME is designed to be a fast and efficient protocol to parse, avoiding to have to scan the entire message to locate boundaries. The length of the attached files is encoded in the message header instead, enabling large attachments to be processed in chunks. The DATA field of an DIME record can contain up to 4 GB of data. Although this is a physical limitation on the amount of data in a single DIME record, there is no limit to the number of records in a DIME message. Since large attachments can be chunked, the DIME specification places no actual limit on the size of attachments. While DIME provided a more efficient processing model it still do not provide an XML Infoset model for the message and attachment. 
As for MIME, DIME breaks the Web services model resulting in, among other
things, that attachments cannot be secured using WS-Security.

\begin{figure}[htp]
 \includegraphics[width=4.6in]{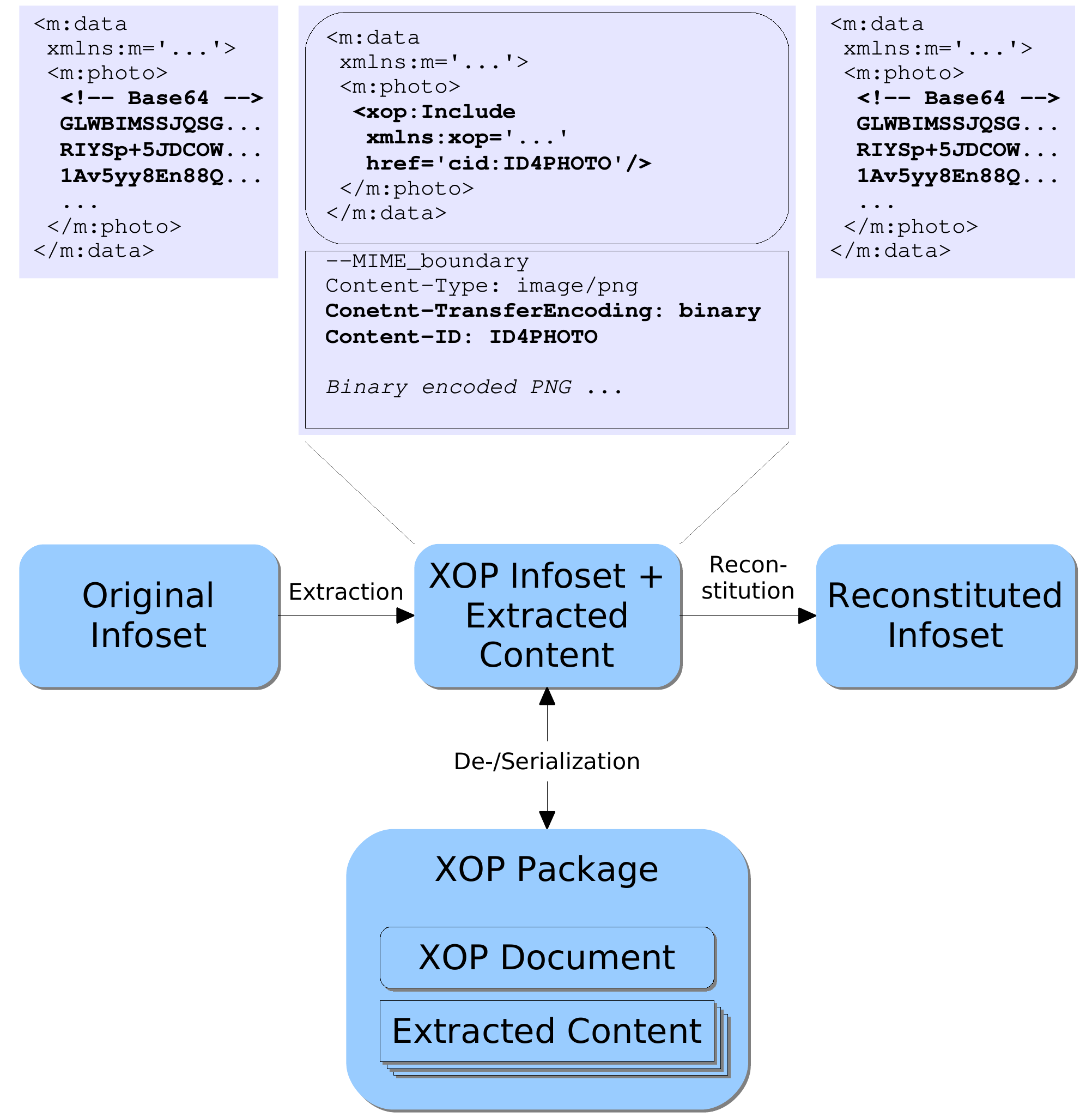}
 \caption{XOP Processing Model}
  \label{fig:xopproc}
\end{figure}

The W3C released a recommendation of a convention called XML-binary Optimized 
Packaging (XOP) \cite{XOP} to provide a way to package XML documents for 
purposes of serialization and transmission. In brief, XOP specifies a method 
for serializing XML Infosets with non-XML, base64-encoded content into MIME 
packages. In the serialization step, an XML document is placed inside a 
so-called XOP package (see Fig.~\ref{fig:xopproc}). Any portions of the 
XML document that are base64-encoded are extracted and optimized. Each 
extracted and optimized chunk is replaced by an \texttt{xop:Include} element 
which refers to the corresponding new location in the XOP package. Thus, 
XOP enables to include binary data alongside with plain-text XML without 
influencing the XML Infoset, hence allowing to apply for example WS-Security 
to the whole message including all binary content. It furthermore promises to 
result in a much smaller dataset than the equivalent base64-encoded data 
without to worry about managing the binary data either on the encoding or the 
decoding side. 

In an attempt to leverage the advantages of the two techniques described 
above: the ``by value'' and ``by reference'' approaches, the W3C developed the 
SOAP Message Transmission Optimization Mechanism (MTOM) \cite{MTOM}
specification.
Actually, MTOM is a ``by reference'' method, since MTOM attachments are streamed as binary data within a MIME message part. Hence, MTOM messages are valid SwA messages, making it fairly easy to pass MTOM attachments to SwA or receive SwA attachments into an MTOM implementation lowering the cost of supporting MTOM for existing SwA implementations. Most notably is the use of the \texttt{xop:Include} element to reference the binary attachment of the message, which is defined in the XOP recommendation. With the use of this exclusive element the attached binary data logically becomes in-line (``by value'') with the SOAP message even though actually it is attached separately. Hence, MTOM provides a compromise between the MIME model and the Web services model since an XML Infoset representation is available.

The paper will focus on MTOM for data transfer with SOAP, since it supports the
efficient data transfer mechanisms specified in the MIME standard and at the
same time allows the application of WS-Security for realizing security
services. Moreover, MTOM is the first data transfer mechanism that is
supported by all major WS platforms. The emphasis is drawn to the signature of
SOAP messages and MTOM attachments respectively, because the 
conventional approach introduces delays at the sending side as will be 
illustrated in the following section.

%%%%%%%%%%%%%%%%%%%%%%%%%%%%%%%%%%%%%%%%%%%%%%%%%%%%%%%%%%%%%%%%%%%%%%%%%%%%%%
\section{Signature of MTOM-optimized SOAP Messages}
The WS-Security standard specifies, that only data within the SOAP enveloped 
should be processed with the defined security mechanisms. Thus, WS-Security 
cannot be applied to SwA or DIME messages directly, but can be used to 
secure MTOM-optimized SOAP messages. (As noted above, there is a separate
WS-Security Profile which can be applied to SwA messages, but it suffers from
the same problems encountered when applying WS-Security to the 
MTOM-optimized SOAP message.)

\begin{figure}[htp]
 \includegraphics[width=4.6in]{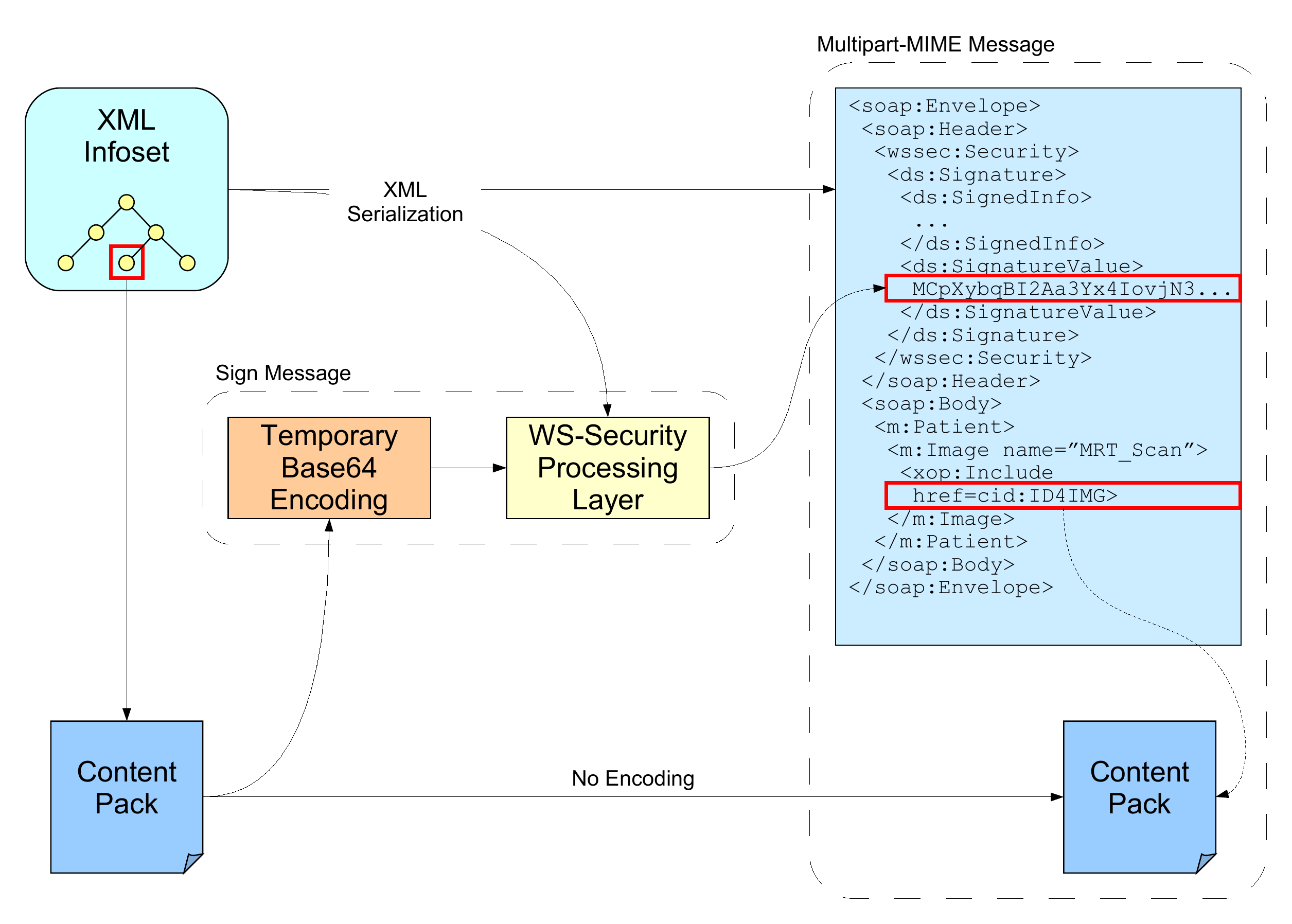}
 \caption{Signature of an MTOM-optimized SOAP Message}
 \label{fig:blcksig}
\end{figure}

With MTOM, everything is inside the SOAP envelope, at least logically. The physical treatment within the endpoints and on the wire is different. Here large (binary) data are handled outside the SOAP envelope to reduce the memory usage and the required amount of data for transmission. Whenever the SOAP message or parts of it containing logically included data have to be processed, the externally managed data becomes temporary part of the message in order to perform the processing. This can be illustrated conveniently by referring to the process of signature generation (see Fig.~\ref{fig:blcksig}).

Before handing the message or parts of it to the WS-Security processing layer, the externally managed content needs to be included. This requires a base64-encoding step, to temporarily construct the XML Infoset. On this volatile XML document the mechanisms defined in WS-Security can be applied and in the context of this paper especially the XML-Signature \cite{XMLSig} processing layer. The outputs of the signature generation process - the digest and signature value - can then be placed into the WS-Security header. Hereafter, the temporarily created message is discard and the content is still managed external to the message in binary format.

%%%%%%%%%%%%%%%%%%%%%%%%%%%%%%%%%%%%%%%%%%%%%%%%%%%%%%%%%%%%%%%%%%%%%%%%%%%%%%
\section{Non-Blocking Signature Approach}
The standard approach to signing a SOAP message when MTOM is used, is to re-create the original XML Infoset before signing the message. As the SOAP envelope is commonly in the first part of multi-part MIME message, the signature must be completed in order to construct the WS-Security header, which for large files this can be a considerable bottleneck.  

\begin{center}
\begin{figure}[htp]
 \includegraphics[width=4.6in]{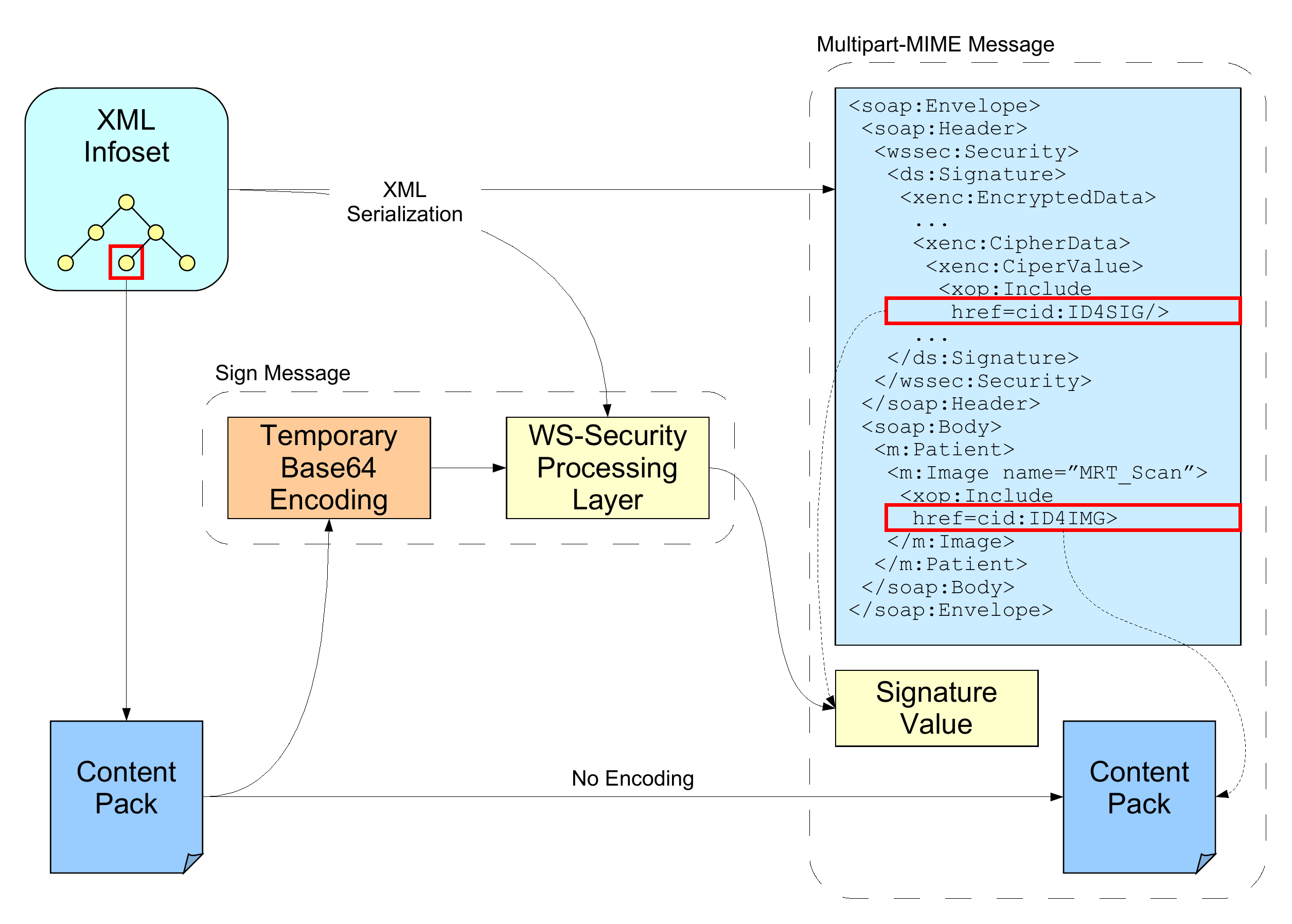}
 \caption{Non-Blocking Signature of an MTOM-optimized SOAP Message}
 \label{fig:nonblcksig}
\end{figure}
\end{center}

To optimized the signature process a non-blocking approach is required.
Ideally, such an approach should calculate the signature while the message is
being sent, i.e., such an approach would be streaming while at the same time
compliant with the current WS-Security specifications and compatible with the
standard SOAP processing model. To realize this ideal approach, the basic idea
introduced in this paper is to include a reference into the
\texttt{ds:Signature} element of the WS-Security header which refers to the
actual signature value contained and send it as the last attachment in the 
multipart MIME format used to convey the message 
(see Fig.~\ref{fig:nonblcksig}). A specific realization of this idea advocated 
here is to use XOP to extract the contents of the \texttt{ds:Signature}, then 
apply MTOM to send this content as the last part of a multi-part MIME message. 
The digest values of the initial parts of the MIME message can be calculated 
while the data is being streamed over the network, leaving the actual 
construction of the signature until the system is ready to send the last part
of the MIME message (see Fig.~\ref{fig:nonblcksig}).

Strictly speaking, XOP may only be applied to content of type \texttt{xsd:base64Binary}; hence, if it is desired to be strictly standard's compliant, one can first encrypt the \texttt{ds:Signature} element, then apply XOP to the \texttt{xenc:CipherValue} element. (This is only a mild restriction since in practice, signing is quite often followed by encrypting.)

Although it would be straight forward to implement this approach from the bottom-up, the purpose of this paper is to explore how it can be made to fit within a standard WS framework, which for concreteness, we take to be the JAX-WS framework. (Although few WS frameworks are fully compliant to the JAX-WS standard, most Java-based frameworks like Axis do adhere to the JAX-WS processing model.) Following the JAX-WS standard, one would use the Handler framework for implementing the WS-Security functionality and the Java Activation Framework (JAF) along with JavaMail for implementing the MTOM functionality. Within JAF, the data is put on the wire through the \texttt{writeTo(OutputStream outputStream)} method of the \texttt{javax.activation.DataHandler}. In particular, there will be one instance of this class for each \texttt{xop:Include} element, with each instance writing to a separate part of the multi-part MIME message. Hence, in order to realize the goal of calculating the digest value while the MTOM data is being sent, this class must be sub-classed and the \texttt{writeTo(OutputStream outputStream)} method overridden to include calculation of the digest while the data is being streamed out. (At this point one could also encrypt the data using key material passed in from a suitable WS-Security Handler.)

Unlike standard implementations where the signature is computed and inserted into the SOAP header within a suitable Handler class, the WS-Security Handler class developed here is only responsible for collecting the security material and passing it to the data handlers as well as inserting the XOP optimized \texttt{ds:Signature} element into the security header. The actual work of preparing the signature is 
delegated to a second class extending \texttt{javax.activation.DataHandler} which is responsible for generating the content of the \texttt{ds:Signature} element and inserting it (or its encrypted counterpart) as the last part of the multi-part MIME message.

%%%%%%%%%%%%%%%%%%%%%%%%%%%%%%%%%%%%%%%%%%%%%%%%%%%%%%%%%%%%%%%%%%%%%%%%%%%%%%
\section{Evaluation and Discussion}
In this section we describe an implementation of the approach discussed in the
previous section. Two popular Java based WS frameworks are investigated here.
The first, Axis2 from Apache has a WS-Security framework in which the MTOM
optimized parts are signed using the approach of reconstructing the original
XML infoset. The second, XFire from Codehouse, does not sign the MTOM
attachments, instead only the elements appearing in the envelop are available
for signing. As XFire does not have a complete WS-Security framework, the
non-blocking approach was implemented in it and compared with the standard
approach from Apache Axis2.

The experimental set-up consisted of an XFire (or Axis) client on one PC and an Apache Tomcat Server hosting the corresponding service on a second PC connected by a $100 \textrm{ Mbps}$ network. The client machined contained an Intel Pentium 4, 3.2 GHz cpu, while the server machine contained a dual AMD Opteron 2.6 GHz cpu.

\begin{center}
\begin{figure}[htbp]
 \includegraphics[width=4.60in]{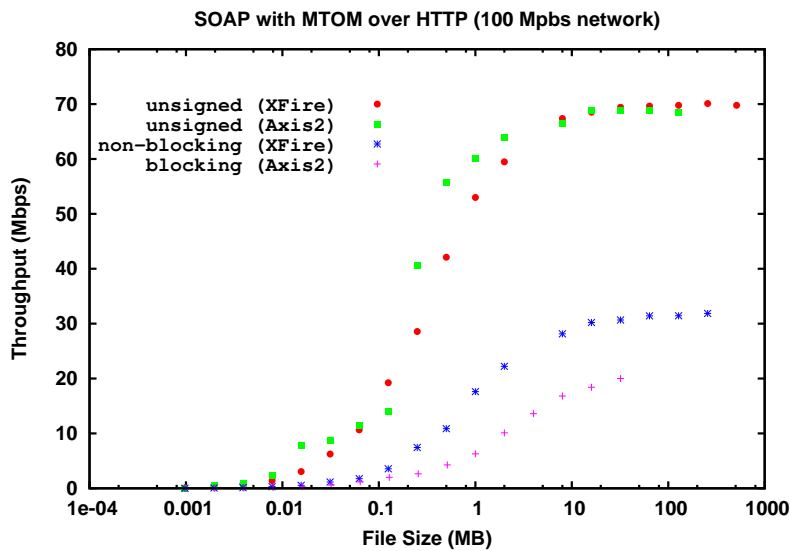}
 \caption{Performance of SOAP Message Signing Approaches}
 \label{fig:perform}
\end{figure}
\end{center}

As a first step, the performance of both frameworks in the absence of any security overhead was measured in order to set the scale for the absolute performance. The results depicted in  Fig.~\ref{fig:perform} show how both frameworks, in the absence of security, are capable of transferring large files with a reasonable efficiency, i.e., the throughput is 70\% of the peak bandwidth. The similarity between the results is to be expected as both frameworks are using the same components at the transport level, namely, the Jakarta Commons HttpClient on the client side and the Apache Tomcat on the server side. Hence, the upper curves in Fig.~\ref{fig:perform} simply demonstrate the efficiency by which large files can be transferred using SOAP/HTTP.

The lower curves show the performance when signing large messages using the the blocking and non-blocking approaches. (For the non-blocking approach, the signature, along with the rest of the message, was encrypted in order to be strictly compliant to the MTOM standard.) As can be seen the non-blocking approach is 50\% faster than the blocking approach, although both approaches are significantly slower than without signature. For the Axis2 framework there is an additional problem in that the JVM crashes with an out-of-memory error when signing large files. Presumably this indicates that Axis2 is trying to completely recreate the original XML infoset in memory before signing. 

Finally, it should be emphasized again, that the present approach and in
particular its specific implementation has been design to enable a smooth and 
seamless integration into existing WS frameworks and to be in conformance 
with the related standards. A more straightforward realization would have 
been to send the SOAP envelope as the last part of the multipart MIME message 
and to set the final part to be the root part. To implement such an approach,
however, the underlying HTTP infrastructure as well as the interface between 
the WS and the HTTP layer would need to support this feature, which at present
they do not.

%%%%%%%%%%%%%%%%%%%%%%%%%%%%%%%%%%%%%%%%%%%%%%%%%%%%%%%%%%%%%%%%%%%%%%%%%%%%%%
\section{Conclusion}
Data services are a basic functionality in service-oriented environments such as Grids. Depending on the concrete application field, integrated security mechanisms are a vital prerequisite. Through the lack of appropriate standards, the application of WS-Security and related specifications has not been easily possible. Starting with MTOM, SOAP messages can be processed with WS-Security. However, when transferring large data sets such as given by medical images, the application of the standard approach for signature generation based on WS-Security to MTOM-optimized SOAP messages introduces a considerable bottleneck. The proposed and presented non-blocking signature generation approach reduces the total transmission time significantly which results in an increased throughput of 50\% in comparision to the standard blocking approach.

%%%%%%%%%%%%%%%%%%%%%%%%%%%%%%%%%%%%%%%%%%%%%%%%%%%%%%%%%%%%%%%%%%%%%%%%%%%%%%
\noindent{\large\bf Acknowledgments}
The authors would like to thank Jun Wang for many fruitful discussions.

%%%%%%%%%%%%%%%%%%%%%%%%%%%%%%%%%%%%%%%%%%%%%%%%%%%%%%%%%%%%%%%%%%%%%%%%%%%%%%

\end{document}